\documentclass[prb,
showpacs,twocolumn,amsmath,amssymb,floatfix,superscriptaddress]{revtex4-2}
\usepackage{graphicx} 
\usepackage{bm}
\usepackage{bbm}
\usepackage{color}
\usepackage{amsmath}
\usepackage{amssymb}
\usepackage{color}
\usepackage{footnote}
\usepackage{comment}
\usepackage{hyperref}
\usepackage{subfigure}
\usepackage{appendix}
\usepackage[applemac]{inputenc}
\usepackage{ulem}

\DeclareRobustCommand{\lcrev}[1]{\textcolor{blue}{#1}}

\begin{document}
\title{Perfect spin nonreciprocity in gated  superconducting altermagnetic heterostructures}

 \author{Pei-Hao Fu}
\email{phy.phfu@gmail.com}
\affiliation{Department of Physics and Astronomy, University of Florence, I-50019 Sesto Fiorentino, Italy}

\author{Jun-Feng Liu}
\affiliation{School of Physics and Materials Science, Guangzhou University, Guangzhou 510006, China}

\author{Luca Chirolli}
\affiliation{Department of Physics and Astronomy, University of Florence, I-50019 Sesto Fiorentino, Italy}
 
\author{Jorge Cayao}
\email{jorge.cayao@physics.uu.se}
\affiliation{Department of Physics and Astronomy, Uppsala University, Box 516, S-751 20 Uppsala, Sweden}
 
\date{\today} 
\begin{abstract}
We consider a superconducting altermagnet heterostructure and demonstrate that the interplay between altermagnetism and a selective filter of transverse momentum channels enables perfect nonreciprocal spin-polarized currents. 
We demonstrate that this nonreciprocity manifests in both local and nonlocal spin currents, signalling the emergence of directionally selective local and nonlocal spin behaviors. 
We show that the selective filter of transverse momentum channels is realized by gating a finite normal region between the superconducting altermagnet and the metallic reservoir, which then directionally selects transport channels that match the momentum-dependent spin-split superconducting altermagnetic states, allowing for nonreciprocal spin-polarized currents. 
We discover that the local and nonlocal spin nonreciprocity features a highly tunable polarity and nearly perfect quality factors, respectively, which is achieved by means of gate voltages and by varying the length of the finite region. 
Moreover, we find that local and nonlocal charge currents also develop a nonreciprocal behavior, whose quality factors can also reach perfect values. 
In all cases, the spin and charge currents are sensitive to variations of the altermagnetic field, a functional dependence that can be exploited to identify the type of altermagnetism. 
Our findings put forward an electrically controllable route towards nonreciprocal superconducting spintronic devices based on altermagnets. 
\end{abstract}
\maketitle

 
\section*{introduction}
The recent discovery of altermagnetism has opened new opportunities for emergent superconducting phases without requiring finite net magnetization
\cite{Yuri2025Superconducting,liu2025review}. This is in great part because altermagnets (AMs) support spin-split Fermi surfaces as in ferromagnets \cite{Brando2016Metallic, Huang2020Recent}, yet exhibit vanishing net magnetization and stray fields due to compensated magnetic ordering, similar to antiferromagnets \cite{Hirohata2023Antiferromagnetic}.
The spin splitting in AMs is of nonrelativistic origin and exhibits an anisotropic behavior,  leading to spin-polarized Fermi surfaces with spin-degenerate nodes depending on the type of AM
\cite{noda2016momentum, NakaNatCommun2019,Hayami19,Ahn2019,Yuanprb20,LiborSAv,NakaPRB2020,Yuanprm21,LiborPRX22,landscape22,MazinPRX22}. These intriguing spin properties of AMs have been shown to be the key for realizing several novel superconducting phases \cite{Yuri2025Superconducting,liu2025review}, 
such as anomalous Josephson effects \cite{zhang2024finite,Ouassou23, yang2025topological,Bo2024,Cheng24,Fukaya2025Josephson,Sun2025Tunable,Zhao2025Orientation,mj4b-2fnr,Pal2025Josephson}, spin-triplet superconductivity \cite{Maeda2025Classification,Chakraborty2024Constraints,khodas2025strain,PhysRevB.111.054520,parshukov2025,mazin2025notes,Fu2025Light,Fu2025Floquet,Yokoyama25floquet,Mukasa2025FiniteMomentum,heinsdorf2025,monkman2025perscurrent}, and nonreciprocal currents \cite{chengdiode24,Sim2024PairDensity,Banerjee2024a,Jiang_2025,Chakraborty25,sharma2025diode,debnath2025,Ruthvik2025FieldFree,sharma2026pUM,Debnath2026Spin,Esin2026Josephson,mondal2026NJDE}; see also Ref.\,\cite{Shaffer2025Theories}.
In this regard, nonreciprocal currents are particularly important because they define elements with a direction-dependent current flow known as diodes \cite{Nagaosa2024Nonreciprocal, Tokura2018Nonreciprocal, Shaffer2025Theories, Nadeem2023Superconducting, Wakatsuki2017Nonreciprocal}, which are the key for superconducting circuits and electronics \cite{braginski2019SC,kochan2025low}. 
To date, nonreciprocity in superconducting AMs has been reported mainly in charge currents but requiring extra ingredients such as spin-orbit coupling or magnetic fields \cite{chengdiode24,Sim2024PairDensity,Banerjee2024a,Jiang_2025,Chakraborty25,sharma2025diode,debnath2025,Ruthvik2025FieldFree,sharma2026pUM,Debnath2026Spin,Esin2026Josephson,mondal2026NJDE}. This not only leaves charge nonreciprocity solely from altermagnetism an open question but it also shows that spin nonreciprocity in superconducting AMs has so far remained seldom studied.

\begin{figure}[!t]
\centering \includegraphics[width=0.48\textwidth]{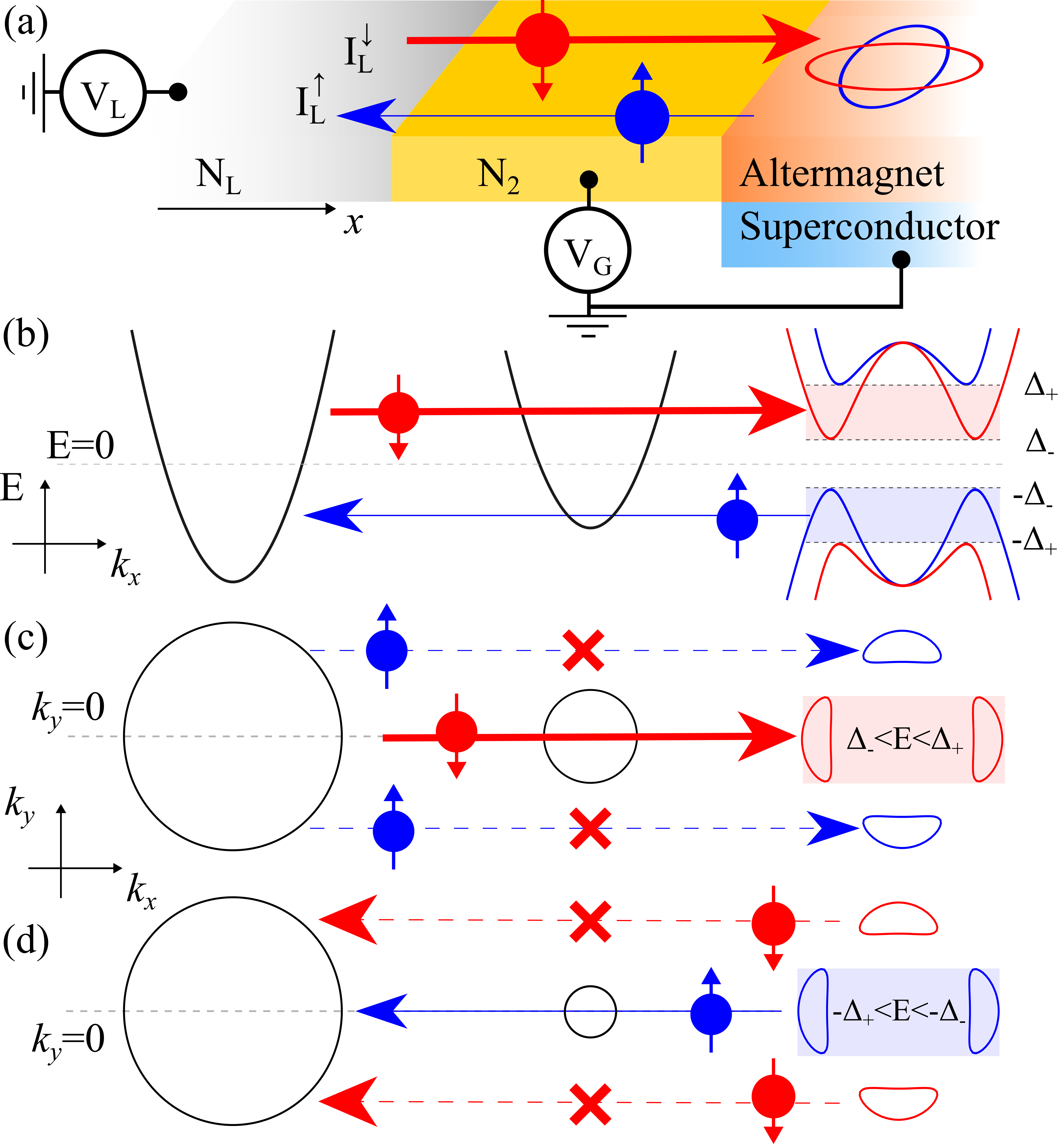}
\caption{
\textbf{Junction geometry,  energy dispersions and contours, and sketch of the spin nonreciprocity.}
(a) Illustration of the studied heterostructure formed by a superconducting altermagnet  (AM) coupled to a   nonaltermagnetic normal region $N_{2}$ (orange) of finite length $d$, which is connected to a left metallic lead (gray). The superconducting altermagnet appears due to an altermagnet (orange) and proximity induced superconductivity from a conventional $s$-wave superconductor (blue). The left lead is subjected to a voltage bias $V_{\rm L}$, while the chemical potential of the normal region $N_{2}$ is controlled by a gate voltage $V_{\rm G}$.
The red thick solid  (blue thin) arrow denotes the current  $I^{\downarrow}_{\text{L}}$ ($I^{\uparrow}_{\text{L}}$) predominantly carried by spin-$\downarrow$ ($\uparrow$) electrons. (b) Energy dispersions in the three regions of the junction  at $k_y=0$, where the shaded regions in the rightmost panel indicate the partially gapped regions in the superconducting AM. (c) Equal-energy contours in the $k_x$-$k_y$ plane for the three regions of the junction at $E>0$, while (d) the same at $E<0$; here, $E$ lies within the red/blue regions in the rightmost panel of (b), which are partially gapped regions of the superconducting AM. The black curves in (b-d) denote spin-degenerate bands in N$_{\text{L}}$ and N$_2$, while blue (red) curves in the rightmost panel indicate the spin split energy bands in the superconducting AM, where quasiparticle states are formed by spin-$\uparrow$ ($\downarrow$) electrons paired with spin-$\downarrow$ ($\uparrow$) holes.   Solid (dashed) arrows in (c,d) denote allowed (blocked) transport channels. 
}
\label{figure1}
\end{figure}

In this work, we consider heterostructures formed by a superconducting AM and a finite 
normal region [Fig.\,\ref{figure1}(a) and Fig.\,\ref{figure4}(a)], and demonstrate that it is possible to realize perfect spin nonreciprocity at vanishing net magnetization and stray fields.
In particular, we focus on $d$-wave AMs with spin-singlet $s$-wave superconductivity, which can be induced by the proximity effect.  We show that gating the finite confined normal region acts as a momentum filter that directionally selects transverse modes matching the spin-split states in the superconducting AM, thereby enabling local and nonlocal nonreciprocal spin-polarized currents.
We find that the nonreciprocal spin currents can be controlled by a gate voltage and the length of the finite confined region, thereby notably permitting the achievement of perfect spin nonreciprocity and a highly tunable polarity, in both local and nonlocal transport. 
For instance, a finite (vanishing) forward (backward) current is generated under positive (negative) bias and is predominantly carried by spin-up (spin-down) electrons, as shown in Figs.\,\ref{figure1}(a) and \ref{figure4}(a) for the local and nonlocal currents, respectively. 
Besides spin currents, we also obtain that charge currents are nonreciprocal and acquire perfect quality factor. 
Yet another consequence of our results is that the spin and charge currents exhibit signatures that can help identify the type and strength of altermagnetism. 
Our results offer a way for spin and charge nonreciprocity based on superconducting AMs, which can be useful for superconducting spintronic with vanishing magnetization.


\section*{Results}


\subsection*{Intrinsic asymmetric spin splitting in superconducting AMs}
We start by showing that the spin splitting of AMs exhibits an intrinsic asymmetry, which is useful for spin nonreciprocity. 
For pedagogical purposes but without loss of generality, we focus on a $d$-wave AM with conventional spin-singlet $s$-wave superconductivity, modelled by an effective Bogoliubov-de Gennes Hamiltonian in the basis of $(c_{k,\uparrow}, c_{k,\downarrow},c^\dag_{-k,\uparrow}, c^\dag_{-k,\downarrow})^{\rm T}$, which reads
\begin{equation}
\label{eq_hqbdg}
\mathcal{H}_{\bm{k}} = \xi_{\bm{k}} \tau_z - \Delta\, \sigma_y \tau_y + M_{\bm{k}} \sigma_z \tau_z\,.
\end{equation}
Here $\sigma_{j}$ ($\tau_{j}$) denotes the $j$th Pauli matrix in spin (Nambu) space, $\xi_{\bm{k}}=\hbar^{2}k^{2}/(2m)-\mu$ is the kinetic energy with chemical potential $\mu$, $\bm{k}=(k_x,k_y)$ is the crystal momentum, $\Delta$ is the spin-singlet $s$-wave pair potential expected to be induced by proximity effect \cite{Buzdin2005Proximity, Yuri2025Superconducting, Cayao2020odd,tanakaReview2024}, and $M_{\bm{k}}$ represents the $d$-wave altermagnetic field.
Explicitly, $M_{\bm{k}}$ takes the form \cite{Yuri2025Superconducting,Smejkal2022a,Smejkal2022}
\begin{equation}
M_{\bm{k}} = J (k/k_{\rm F})^2 \cos\!\left(2\theta_k - 2\theta_J\right),
\label{eq_mk}
\end{equation}
where $k_{\rm F}=\sqrt{2m\mu}/\hbar$ is the Fermi wave vector, $J$ denotes the altermagnetic strength, $\theta_J$ specifies the orientation of the altermagnetic axis relative to the $k_x$ direction, and $\theta_k=\arctan(k_y/k_x)$ denotes crystalline momentum direction. Thus, for $\theta_J=0$ ($\theta_J=\pi/4$), Eq.\,(\ref{eq_mk}) describes a $d_{x^2-y^2}$-wave ($d_{xy}$-wave) AM. 

The altermagnetic spin splitting is directly revealed in the quasiparticle spectrum of Eq.\,(\ref{eq_hqbdg}), which reads
\begin{equation}
E_{\bm{k}}^{\gamma ,\beta} = \gamma M_{\bm{k}} + \beta \sqrt{\xi_{\bm{k}}^{2} + \Delta^{2}}\,,
\label{eq_Eqbdg}
\end{equation}
where $\beta=\pm$ labels the electron- and hole-like quasiparticle branches.
The index $\gamma=+\,(-)$ denotes the sector of Eq.\,(\ref{eq_hqbdg}) with spin-$\uparrow$ ($\downarrow$) electrons pairing with spin-$\downarrow$ ($\uparrow$) holes. The first term in Eq.\,(\ref{eq_Eqbdg}) shows that a finite altermagnetic field $M_{\bm{k}}$ [Eq.\,(\ref{eq_mk})] generates the spin splitting, resulting in $E_{\bm{k}}^{+ ,\beta}\neq E_{\bm{k}}^{- ,\beta}$.

The spin splitting manifests itself both in energy and momentum space.
The rightmost panel of Fig.\,\ref{figure1}(b) schematically shows the dispersion of Eq.\,(\ref{eq_Eqbdg}) as a function of $k_x$ for $\theta_J=0$ and $k_y=0$ ($\theta_k=0$).
In this case, $M_{\bm{k}}>0$ for $k_x\neq0$, which shifts the branches with $\gamma=+$ ($\gamma=-$) upward (downward), producing the energy splitting between opposite spin sectors.
Furthermore, the altermagnetic field $M_{\bm{k}}$ induces anisotropic spin splitting that depends on the crystalline momentum direction $\theta_k$.
Specifically, $M_{\bm{k}}$ reaches positive (negative) extrema at $\theta_k=\theta_J+n\pi$ [$\theta_k=\theta_J+(2n+1)\pi/2$] and vanishes at $\theta_k=\theta_J+(2n+1)\pi/4$, where $n\in\mathbb{Z}$.
Fig.\,\ref{figure1}(b) is the case with $\theta_J=\theta_k=0$ having maximal $M_{\bm{k}}$. Hence, $M_{\bm{k}}$ generates an anisotropic spin-dependent spectrum in $d$-wave AMs that depends strongly on the momentum direction.

Further insight into the anisotropic spin splitting can be obtained from the band edges of Eq.\,(\ref{eq_Eqbdg}), which occur at $k=k_{\rm F}$ where the superconducting gap opens and are given by
\begin{equation}
E_{\text{edge}}^{\gamma,\beta}(\theta_k) = \gamma M_{k_{\rm F},\theta_k} + \beta \Delta\,.
\label{eq_eg}
\end{equation}
Here $M_{k_{\rm F},\theta_k}=J\cos(2\theta_J-2\theta_k)$ is the altermagnetic field given by Eq.\,(\ref{eq_mk}) with $k\rightarrow k_F$ while retaining the angular dependence on $\theta_k$.
For $\theta_k=\theta_J+(2n+1)\pi/4$, one has $M_{k_{\rm F},\theta_k}=0$, which yields spin-degenerate band edges at $\pm\Delta$.
In contrast, at $\theta_k=\theta_J+n\pi/2$ the field $M_{k_{\rm F},\theta_k}$ is maximal and determines the upper and lower bounds of the positive-energy band edges,
\begin{equation}
E_{\text{edge}}^{\gamma,+}(\theta_J+n\pi)\equiv\Delta_\gamma
= \left|\Delta + \gamma J\right|\,,
\label{eq_Deltapm}
\end{equation}
which are indicated by the short dashed gray horizontal lines in the rightmost panel of Fig.\,\ref{figure1}(b).
These correspond to the highest ($\gamma=+$) and lowest ($\gamma=-$) band edges at positive energies, with symmetric counterparts at negative energies. Thus, Eq.\,(\ref{eq_Deltapm}) defines three distinct energy regimes in the quasiparticle spectrum, as indicated  by the shaded areas in the rightmost panel of Fig.\,\ref{figure1}(b):  (i) For $|E|<\Delta_-$, the system is fully gapped without quasiparticle excitations.
(ii) For $\Delta_-<|E|<\Delta_+$, the gap becomes direction dependent and quasiparticle states emerge in opposite spin sectors:
In particular, for $E>0$, quasiparticle states around $k_y=0$ ($k_x=0$) are dominated by the spin-$\uparrow$ ($\downarrow$) sector; for $E<0$, the spin character is reversed, see Figs.\,\ref{figure1}(c) and \ref{figure1}(d).
(iii) For $|E|>\Delta_+$, the spectrum becomes gapless with quasiparticles occupying all momentum directions.

As a result, the characteristic gaps $\Delta_\pm$ and  band edges $E_{\text{edge}}^{\gamma,\beta}$ encode the energy- and momentum-dependent and asymmetric spin splitting of the quasiparticle spectrum arising from the interplay between $d$-wave altermagnetism and conventional superconductivity.
The resulting asymmetric spectrum with respect to $E=0$ in each spin sector 
, akin to what occurs in ferromagnetic junctions \cite{Buzdin2005Proximity,Giazotto2008Superconductors},  has been experimentally identified as a key ingredient for nonreciprocal transport \cite{Strambini2022Superconducting,DeAraujo2024Superconducting,PhysRevApplied.4.044016,Geng2023Superconductor,Putilov2024Nonreciprocal}. 
Despite this similarity, the anisotropic and asymmetric spin splitting due to altermagnetism is a distinctive feature of AM-superconductor hybrids without a   ferromagnetic analog.  As we show below, these spin asymmetries are useful for realizing spin-polarized transport in junctions based on superconducting AMs, as illustrated in Fig.\,\ref{figure1}(a).


\begin{figure*}[t]
\centering \includegraphics[width=0.99\textwidth]{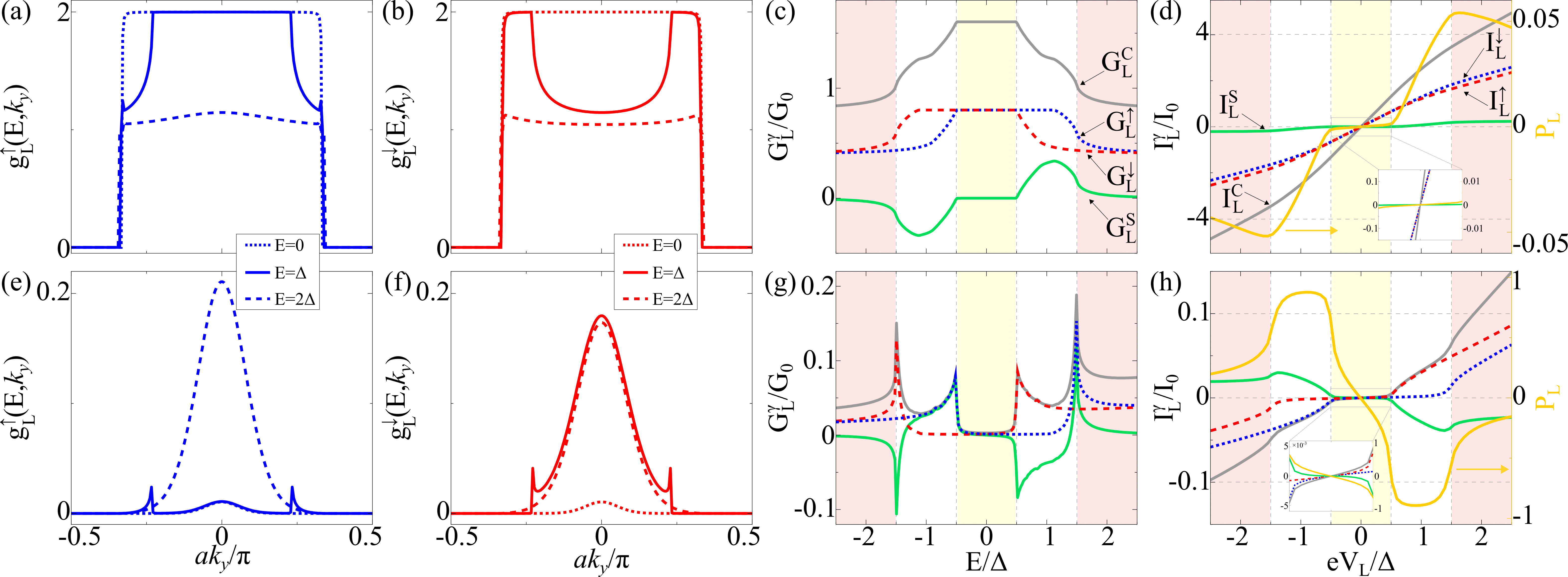}
\caption{
\textbf{Gate-controlled local conductance and currents.}
(a-d) Local conductance and currents at  $eV_{\rm G}=\mu=50\Delta$ in a junction    having $d_{x^{2}-y^{2}}$-wave altermagnetism. (a,b)  Energy- and momentum-resolved dimensionless conductance  $g_{\rm L}^{\sigma}$ for incident spin-$\uparrow$ and spin-$\downarrow$ electrons, respectively. (c) Energy-resolved local spectral conductances  $G_{\rm L}^{\gamma}$ as a function of $E$. (d) Local spin and charge currents $I_{\rm L}^{\gamma}$ versus $eV_{\rm L}$, with the right $y$ axis showing the spin polarization $P_{\rm L}$.  In (c,d), $\gamma \in \{\uparrow,\downarrow,\text{C},\text{S}\}$ denotes the spin-$\uparrow$, spin-$\downarrow$, charge, and spin conductances as well as the respective currents. The yellow shaded, unshaded, and red shaded regions in (c,d) correspond to the gapped, partially gapped, and gapless regimes of the BdG spectrum defined by Eq.\,(\ref{eq_Deltapm}).  (e-h): The same as (a-d) but with $eV_{\rm G}=0.5\Delta$. Parameters: $\mu=t$, $\Delta=0.02\mu$, $J=0.5\Delta$, $\theta_{\rm J}=0$, $d=5a$, $t_c=1$, $\kappa_{\rm B} T=0.01\Delta$, $a=1$, $t=1$, and   $\Delta_\pm=(1\pm0.5)\Delta$.
}
\label{figure2}
\end{figure*}

\subsection*{Gate-controlled local spin-polarized currents}
After showing that superconducting AMs harbor an intrinsic asymmetric spin splitting under $d_{x^{2}-y^{2}}$-wave altermagnetism, in this part, we explore how it can be exploited for obtaining spin-polarized currents. 
While the altermagnetic spin splitting might be seen to be sufficient [right panel of Fig.\,\ref{figure1}(b)], realizing spin-polarized transport is not a trivial task. 
This is because both spin channels contribute to transport, leading to compensated spin currents [see rightmost panels of Figs.\,\ref{figure1}(c,d)]. However, owing to the momentum-dependent spin splitting in AMs, spin-polarized transport can be achieved by selecting transverse momentum channels, while maintaining the altermagnetic properties. 
To implement such a momentum selection mechanism, we insert a nonaltermagnetic normal region of finite length (N$_2$) between the superconducting AM and the normal metallic reservoir (N$_{\rm L}$) needed for transport, resulting in the heterostructure shown in Fig.\,\ref{figure1}(a). 
The finite normal region N$_2$ has a Fermi surface that can be controlled by means of a gate voltage via its chemical potential. 
This makes $N_{2}$  as a momentum filter that selects transverse modes with propagation directions almost aligned with the junction.
Therefore, the transmitted modes match the spin-split states in the superconducting AM, enabling spin-polarized currents [see Fig.\,\ref{figure1}(b-d)].
In the following, we discuss  the calculation of current flowing through the device shown in Fig.\,\ref{figure1}(a) and the effect of gating for realizing spin polarization and nonreciprocity.

\textbf{Current through the device of Fig.\,\ref{figure1}(a).} To properly inspect the momentum filtering effect and the resulting spin transport, we model the junction in Fig.\,\ref{figure1}(a) by discretizing Eq.\,(\ref{eq_hqbdg}) on a lattice along the $x$ direction, while keeping $k_y$ as a good quantum number.
The nearest-neighbor hopping amplitude is $t=\hbar^2/(2ma^2)$, where $a$ is the lattice constant. 
Details of the junction model are provided in the Methods section. 
In the following, for simplicity we set $t=1$ as the energy unit and assume identical hopping in all regions.
The chemical potentials in N$_{\text{L}}$ and the superconducting region are taken to be equal to $\mu\gg\{\Delta,J\}$, such that transport is primarily controlled by the gated region and the interface transparency, characterized by the dimensionless parameter $t_c/t$. 
The gate voltage in the N$_2$ region is the key ingredient controlling the transport, which shifts the electrostatic potential in the N$_2$ region.
In principle, the electrostatic potential is obtained by solving the BdG-Poisson equation for the NS junction geometry \cite{dominguez2017zero,Antipov2018Effects,Woods2018Effective, Mikkelsen2018Hybridization,Winkler2019Unified,Liu2021Electronic} and may spatially depend on the applied bias voltage \cite{Lesovik1997Nonlinearity, Melo2021Conductance}.
For simplicity, we approximate the gate effect by setting $\mu_{\rm N_2} = eV_{\rm G}$ \cite{Datta2005Quantum}.
This approximation captures the essential momentum filtering induced by the gate [Fig.\,\ref{figure1}(b-d)], and the qualitative results presented below are expected to remain valid when a more detailed electrostatic modeling is included. 

The gate-defined spin-polarized transport in the junction is characterized by the relation between the local current flowing in N$_{\rm L}$ driven by bias $eV_{\text{L}}$, which is  \cite{Blonder1982Transition}
\begin{equation}
I_{\text{L}}^{\sigma}(eV_{\text{L}}) = \frac{1}{e} \int dE\, G_{\text{L}}^{\sigma}(E)
\left[f_{\text{L}}(E,eV_{\text{L}})-f_{\text{S}}(E)\right],
\label{eq_iv}
\end{equation}
where $G_{\text{L}}^{\sigma}(E)$ is the   spectral conductance for incident electrons with spin $\sigma\in\{\uparrow,\downarrow\}$, while 
 $f_{\text{L}}(E,eV_{\text{L}})=f(E-eV_{\text{L}})$ and $f_{\text{S}}(E)=f(E)\equiv [1+e^{E/(\kappa_{\rm B} T)}]^{-1}$ are the Fermi-Dirac distributions in the N$_{\text{L}}$ and S regions, respectively; here $T$ is the temperature and $\kappa_{\rm B}$ is the Boltzmann constant. In Eq.\,(\ref{eq_iv}), $G_{\text{L}}^{\sigma}(E)$ is given by
\begin{equation}
G_{\text{L}}^{\sigma}(E) = G_{0} \sum_{k_{y}} g_{\text{L}}^{\sigma}(E,k_y),
\label{eq_Gx}
\end{equation} 
where $G_0 = e^2 W / (2\pi\hbar)$ is the conductance unit in the sample with width $W$ and
\begin{equation}
g_{\text{L}}^{\sigma}(E,k_y) = 1 - R^{\sigma}_{\text{N}} + R^{\sigma}_{\text{A}},
\label{eq_Ts}
\end{equation}
is the energy- and momentum-resolved dimensionless conductance determined by the normal ($R^{\sigma}_{\text{N}}$) and Andreev reflection ($R^{\sigma}_{\text{A}}$) probabilities. Here, $R^{\sigma}_{\text{N}}$ and $R^{\sigma}_{\text{A}}$
are obtained using lattice Green's functions \cite{LopezSancho1985Nonorthogonal,LopezSancho1984Quick,LopezSancho1985Highly,Anantram2008Modeling,san2013multiple,Fu2022,Fu2022b,Fu2020Transport}, as detailed in the Methods section.
From Eqs.\,(\ref{eq_Gx}) and (\ref{eq_iv}), the charge (C) and spin (S) conductances are found as
\begin{subequations}
\begin{align}
G_{\text{L}}^{\text{C}}(E) &= G_{\text{L}}^{\uparrow}(E) + G_{\text{L}}^{\downarrow}(E), \label{eqGC}\\
G_{\text{L}}^{\text{S}}(E) &= G_{\text{L}}^{\uparrow}(E) - G_{\text{L}}^{\downarrow}(E), \label{eqGS}
\end{align}
\end{subequations}
and the corresponding currents as
\begin{subequations}
\label{spinchargecurrents}
\begin{align}
I_{\text{L}}^{\text{C}}(eV_{\text{L}}) &= I_{\text{L}}^{\uparrow}(eV_{\text{L}}) + I_{\text{L}}^{\downarrow}(eV_{\text{L}}), \label{eqILC}\\
I_{\text{L}}^{\text{S}}(eV_{\text{L}}) &= | I_{\text{L}}^{\uparrow}(eV_{\text{L}}) | - | I_{\text{L}}^{\downarrow}(eV_{\text{L}}) |, \label{eqILS}
\end{align}
\end{subequations}
which are the measurable observables of the junction. 
The spin transport is further characterized by the spin polarization \cite{Buzdin2005Proximity,Giazotto2008Superconductors}, 
\begin{equation}
P_{\text{L}}(eV_{\text{L}})=\frac{I_{\text{L}}^{\text{S}}(eV_{\text{L}})}{|I_{\text{L}}^{\text{C}}(eV_{\text{L}})|},
\label{eqP}
\end{equation}
which is defined by normalizing the spin current to the charge current. 
Here, $P_{\text{L}}>0$ ($<0$) corresponds to spin-$\uparrow$ (spin-$\downarrow$) polarization. 
Perfect spin polarization is achieved when $P_{\text{L}}=\pm1$, indicating that the current is solely carried by spin-$\uparrow$ ($\downarrow$) electrons \cite{Buzdin2005Proximity,Giazotto2008Superconductors}.
In the following, we investigate transport in the junction and the effect of gating by analyzing the spin-resolved conductance $g_{\text{L}}^{\sigma}(E,k_y)$ [Eq.\,(\ref{eq_Ts})] and the resulting conductance and current defined in Eqs.\,(\ref{eq_Gx}) and (\ref{eq_iv}), respectively.

\textbf{Effect of gating on spin polarization.}
To reveal the effect of the gate voltage, in Figs.\,\ref{figure2}(a-d) we first assess transport in a \textit{not} gated case $eV_{\rm G}=\mu$ at distinct energies as a benchmark for comparison with $d_{x^{2}-y^{2}}$-wave altermagnetism. 
In this case, $g_{\text{L}}^{\sigma}(E,k_y)$ versus $k_{y}$ in Figs.\,\ref{figure2}(a,b) show\lcrev{s} that all transverse modes with different $k_y$ from the N$_L$ lead are injected into the superconducting lead. 
The $k_y$ dependence of the spin-resolved conductance $g_{\text{L}}^{\sigma}(E,k_y)$ [Figs.\,\ref{figure2}(a,b)] is therefore determined by the three distinct regimes in the spectrum of the superconducting AM illustrated in Fig.\,\ref{figure1}(b-d). 
In the fully gapped region $|E|<\Delta_-$, the dotted lines in Figs.\,\ref{figure2}(a,b) show that $g_{\text{L}}^{\uparrow}(E,k_y)=g_{\text{L}}^{\downarrow}(E,k_y)=2$ due to perfect Andreev reflections and the absence of normal reflections. 
After summing over all transverse modes [Eq.\,(\ref{eq_Gx})], the total spin-resolved conductances $G_{\text{L}}^{\uparrow}$ and $G_{\text{L}}^{\downarrow}$ are degenerate. The resulting charge conductance exhibits a plateau while the spin conductance vanishes, as shown by the yellow shaded region in Fig.\,\ref{figure2}(c).

Spin-dependent transport emerges in the partially gapped region $\Delta_-<|E|<\Delta_+$, as shown by solid curves in Figs.\,\ref{figure2}(a,b): here, the high conductance state for spin-$\uparrow$ ($\downarrow$) with $g_{\text{L}}^{\uparrow(\downarrow)}=2$ or $k_y$ close to (away from) $0$.
This behavior can be understood from the schematic equal-energy contour shown in Fig.\,\ref{figure1}(c): For $k_y\approx0$, the spin-$\uparrow$ spectrum is gapped, and the Andreev reflection dominates transport, yielding $g_{\text{L}}^{\uparrow}(E,k_y)=2$; in contrast, the spin-$\downarrow$ states are gapless, so that the injected electrons are transmitted into the superconductor, reducing the Andreev reflection and thus the conductance [Figs.\,\ref{figure2}(a,b)].
When $k_y$ moves away from $0$, the momentum-dependent spin splitting reverses in the AM-superconductor hybrid, resulting in reduced (enhanced) conductance for spin-$\uparrow$ (spin-$\downarrow$) electrons. 
As a result, the total conductance obtained from Eq.\,(\ref{eq_Gx}) becomes spin dependent, resulting in a finite spin conductance $G_{\text{L}}^{\text{S}}$ in the unshaded region of Fig.\,\ref{figure2}(c).
For $|E|>\Delta_+$, both spin subbands in the AM-superconductor hybrid become gapless, and the resulting transport in this regime is dominated by single-particle processes that lead to
$g_{\text{L}}^{\uparrow}(E,k_y)\approx g_{\text{L}}^{\downarrow}(E,k_y)\approx1$, see dashed curves in Figs.\,\ref{figure2}(a,b). Thus, spin-resolved conductances $G_{\text{L}}^{\sigma}$ become nearly degenerate, with a negligible $G_{\text{L}}^{\text{S}}$, as shown in the red shaded region of Fig.\,\ref{figure2}(c).

The behavior of the conductances exposed above has a direct effect on the local charge ($I_{\text{L}}^\text{C}$) and spin ($I_{\text{L}}^\text{S}$) currents defined by Eqs.\,(\ref{spinchargecurrents}). 
To unveil this, in Fig.\,\ref{figure2}(d) we present $I_{\text{L}}^\text{C}$ and $I_{\text{L}}^\text{S}$ versus $eV_{\rm L}$ at $eV_{\rm G}=\mu$; therein, we also show the spin-resolved components $I_{\text{L}}^{\sigma}$ and spin polarization $P_{\rm L}$ obtained from Eq.\,(\ref{eq_iv}) and Eq.\,(\ref{eqP}), respectively. 
We observe that, for $|eV_{\text{L}}|<\Delta_-$, the identical spin-resolved currents result in a vanishing spin current ($I_{\text{L}}^{\text{S}}=0$) and spin polarization ($P_{\text{L}}=0$), see the inset of Fig.\,\ref{figure2}(d).
In the partially gapped region, with $\Delta_-<|eV_{\text{L}}|<\Delta_+$, both the spin current $I_{\text{L}}^{\text{S}}$ and the spin polarization $P_{\text{L}}$ become finite; this originates due to the spin-dependent conductance in Fig.\,\ref{figure2}(c), as a result of the momentum-dependent spin-split equal-energy contours in the superconducting AM [Fig.\,\ref{figure1}(a)]. 
In the gapless regime, when $|eV_{\text{L}}|>\Delta_+$, the spin polarization $P_{\text{L}}$ decreases as the spin current saturates, while the charge current continues to increase with the applied bias $eV_{\rm L}$. At this point, we emphasize that, in spite of the spin-splitting due to altermagnetism, the spin polarization $P_{\text{L}}$ is almost negligible, with a maximum value of $P_{\text{L}}\approx5\%$, see right $y$ axis in Fig.\,\ref{figure2}(d). 
This, of course, reflects that, albeit there is an overall spin current due to AMs, the charge current is entirely dominant when the finite region $N_{2}$ is not gated.
We have also verified that qualitatively identical spin and charge current behavior is obtained in the absence of the N$_2$ region, demonstrating that spin-polarized transport is negligible in a gate-free NS junction formed by superconducting AMs.




We now assess the regime with a gated $N_{2}$ region by tuning its gate voltage $V_{\rm G}$ to a value comparable to the superconducting gap, $eV_{\rm G}=0.5\Delta=\mu/100$. 
As already anticipated at the beginning of this section, this regime permits filtering momentum states such that the asymmetric momentum-dependent spin splitting of AMs is fully exploited for inducing strong spin-polarized currents. 
This is demonstrated in Figs.\,\ref{figure2}(e-h), where we present the $k_{y}$ dependence of $g_{\text{L}}^{\sigma}(E,k_y)$ as well as $G_{\text{L}}^{\gamma}(eV_{\rm L})$, $I_{\text{L}}^{\gamma}(eV_{\rm L})$ and $P_{\rm L}(eV_{\rm L})$ versus $eV_{\rm L}$. 
The first observation is that the momentum-resolved conductance $g_{\text{L}}^{\sigma}(E,k_y)$ in Figs.\,\ref{figure2}(e,f) exhibits dependences on $k_{y}$, $E$, and spin $\sigma$ that are stronger than when the finite region $N_{2}$ was not gated in Figs.\,\ref{figure2}(a,b).
Because of the small Fermi surface restricted by the gated $N_{2}$ region [Figs.\,\ref{figure1}(c,d)], the conductances now exhibit peaked profiles with dominant contributions from the transverse modes around $k_y\sim0$, see Figs.\,\ref{figure2}(e,f).
Furthermore, for $|E|<\Delta_-$, the conductances $g_{\text{L}}^{\sigma}(E,k_y)$ gain a spin-dependent suppression, due to enhanced normal reflections, which occur because the small Fermi surface in the N$_2$ region drastically mismatches those in the N$_L$ and the superconducting AM regions.
In contrast, in the gapless regime $|E|>\Delta_+$, conductances $g_{\text{L}}^{\sigma}(E,k_y)$ are enhanced due to the quasiparticle transmission. 
This enhancement is spin-dependent, as revealed by the different values of $g_{\text{L}}^{\uparrow}(E,k_y)$ and $g_{\text{L}}^{\downarrow}(E,k_y)$ at $k_y\sim0$. 

Interestingly, the role of the gate becomes most significant in the partially gapped energy window $\Delta_-<|E|<\Delta_+$, as revealed by the solid curves in Figs.\,\ref{figure2}(e,f): in this regime, the momentum resolved conductance $g_{\text{L}}^{\sigma}(E,k_y)$  for up spin electrons is strongly suppressed, while it remains finite for spin down electrons. 
This peculiar behavior of  $g_{\text{L}}^{\sigma}(E,k_y)$ originates because the gated region selects transverse momenta $k_y$ close to $0$, which match the spin-$\downarrow$ ($\uparrow$) states in the superconducting AM region and allow electrons with spin -$\downarrow$ ($\uparrow$) to flow for $eV_L>0$ ($eV_L<0$), see Fig.\,\ref{figure1}(c,d). 
In contrast, the spin-$\uparrow$ (-$\downarrow$) states require larger $k_y$ values that are blocked by the small equal-energy contour in the gated region.

The striking behavior of $g_{\text{L}}^{\sigma}(E,k_y)$ when $N_{2}$ is gated therefore strongly influences the spectral conductances $G_{\text{L}}^{\gamma}(eV_{\rm L})$ and currents $I_{\text{L}}^{\gamma}(eV_{\rm L})$, as shown in Figs.\,\ref{figure2}(g,h). In fact, the reduced values of $g_{\text{L}}^{\sigma}(E,k_{y})$ for $|E|<\Delta_-$ lead to negligible values of $G_{\text{L}}^{\gamma}(eV_{\rm L})$, see yellow shaded region in Figs.\,\ref{figure2}(g,h), respectively.
We note that, although the currents in this bias window are negligible, they exhibit spin-dependent behavior, giving rise to a finite spin polarization, see the inset of Fig.\,\ref{figure2}(h).
This contrasts with the non-gated case, where the currents are spin degenerate, see inset of Fig.\,\ref{figure2}(d).
Interestingly, after increasing the bias within the partially gapped energy window $\Delta_-<|E|<\Delta_+$, the conductance $g_{\text{L}}^{\sigma}(E,k_y)$ is mostly carried by spin down electrons, making $G_{\text{L}}^{\gamma}(eV_{\rm L})$ and   $I_{\text{L}}^{\gamma}(eV_{\rm L})$ in this region to be almost entirely carried by such a single spin channel, see unshaded regions of Figs.\,\ref{figure2}(g,h). 
This behavior signals spin-polarized transport that is characterized by the spin polarization [Eq.\,(\ref{eqP})] reaching $P_{\text{L}}\sim\pm1$ only in the partially gapped regime, see yellow curve in Fig.\,\ref{figure2}(h). 
By further increasing the bias into the gapless regime $|E|>\Delta_+$, the spin polarization is slightly reduced and saturates due to finite spin-dependent quasiparticle transmission. Thus, by comparing Fig.\,\ref{figure2}(h) with Fig.\,\ref{figure2}(d), the interplay between gating $N_{2}$ and superconducting AM gives rise to spin-polarized currents.

In addition to inducing spin-polarized transport, gating introduces two further measurable consequences.
First, the spectral conductance $G_{\text{L}}^{\gamma}(E)$ provides a 
direct way to measure the altermagnetic strength. In fact, comparing Fig.\,\ref{figure2}(c) and Fig.\,\ref{figure2}(g), we see that reducing the value of $eV_{\rm G}$   suppresses the conductance plateau and produces sharp peaks at the band edges $E_{c1}=\Delta_-=\Delta-J$ and $E_{c2}=\Delta_+=\Delta+J$, marked by vertical gray dashed lines in Fig.\,\ref{figure2}(g). These characteristic energies are independent of the altermagnetic orientation $\theta_J$ [see Eq.\,(\ref{eq_Deltapm})] and therefore provide a direct transport probe of the altermagnetic strength, $J=(E_{c2}-E_{c1})/2$.
It is worth noting that, since the superconducting gap $\Delta$ in the considered superconducting AM   is proximity induced and inevitably   differs from the parent superconductor gap $\Delta'$, the   peaks of $G_{\text{L}}^{\gamma}(E)$ also allow to quantify the  proximity efficiency  as $q=\Delta/\Delta'$, where $\Delta=(E_{c2}+E_{c1})/2$ 
A second effect introduced by the gate   is the emergence of nonzero spin and charge current nonreciprocities, which originate as a result of the spin and charge currents possessing different magnitudes when the bias polarity is reversed. 
This is an emergent diode behavior \cite{Nadeem2023Superconducting,Misaki2021Theory,davydova2022,cayao2024_JD,79tj-c3y4,Chirolli2025Diode,hf7s-f7tj,Fu2022b,yerin2026SDEfractal,Muhammad2026Quantum}, sometimes referred to as rectification \cite{ideue2017,Rosdahl2018Andreev, Zazunov2024Approaching,Fu2025Implementation}, which resembles a transistor \cite{Datta2005Quantum} and can be controlled by the applied gate voltage, as discussed below. 

\subsection*{Local spin and charge nonreciprocity}
To quantify the nonreciprocity of the local spin and charge currents
 $I^{\text{C(S)}}_\text{L}$ driven by $eV_\text{L}$, here we investigate the charge and spin quality factors defined as \cite{Sze2012,ideue2017, Strambini2022Superconducting}
\begin{equation}
Q_{\text{L}}^{\gamma}(eV_{\text{L}})=
\frac{|I_{\text{L}}^{\gamma}(+eV_{\text{L}})|-|I_{\text{L}}^{\gamma}(-eV_{\text{L}})|}
{|I_{\text{L}}^{\gamma}(+eV_{\text{L}})|+|I_{\text{L}}^{\gamma}(-eV_{\text{L}})|}\,,
\label{eqQ}
\end{equation}
where $I^{\gamma}(eV_{\text{L}})$ corresponds to the charge ($\gamma={\rm C}$) and spin ($\gamma={\rm S}$) currents  under bias $V_{\text{L}}$ obtained from Eq.\,(\ref{eq_iv}) and discussed in previous section. The quality factors $Q_{\text{L}}^{\gamma}$ characterize the charge (spin)   nonreciprocal efficiency. In particular, a $Q_{\text{L}}^{\text{C(S)}}\neq0$ indicates that the charge (spin) currents have different magnitudes under reversed biases. 
The limiting values $Q_{\text{L}}^{\gamma}=\pm1$ correspond to perfect nonreciprocity, where a finite current flows in one direction while it vanishes in the opposite direction. 
Also, the sign of $Q_{\text{L}}^{\gamma}$ determines the polarity of the nonreciprocal transport.
Together with the spin polarization $P_{\text{L}}$ [Eq.\,(\ref{eqP})], these quantities characterize the local spin and charge nonreciprocities of the device shown in Fig.\,\ref{figure1}(a).

\begin{figure}[t]
\centering \includegraphics[width=0.49\textwidth]{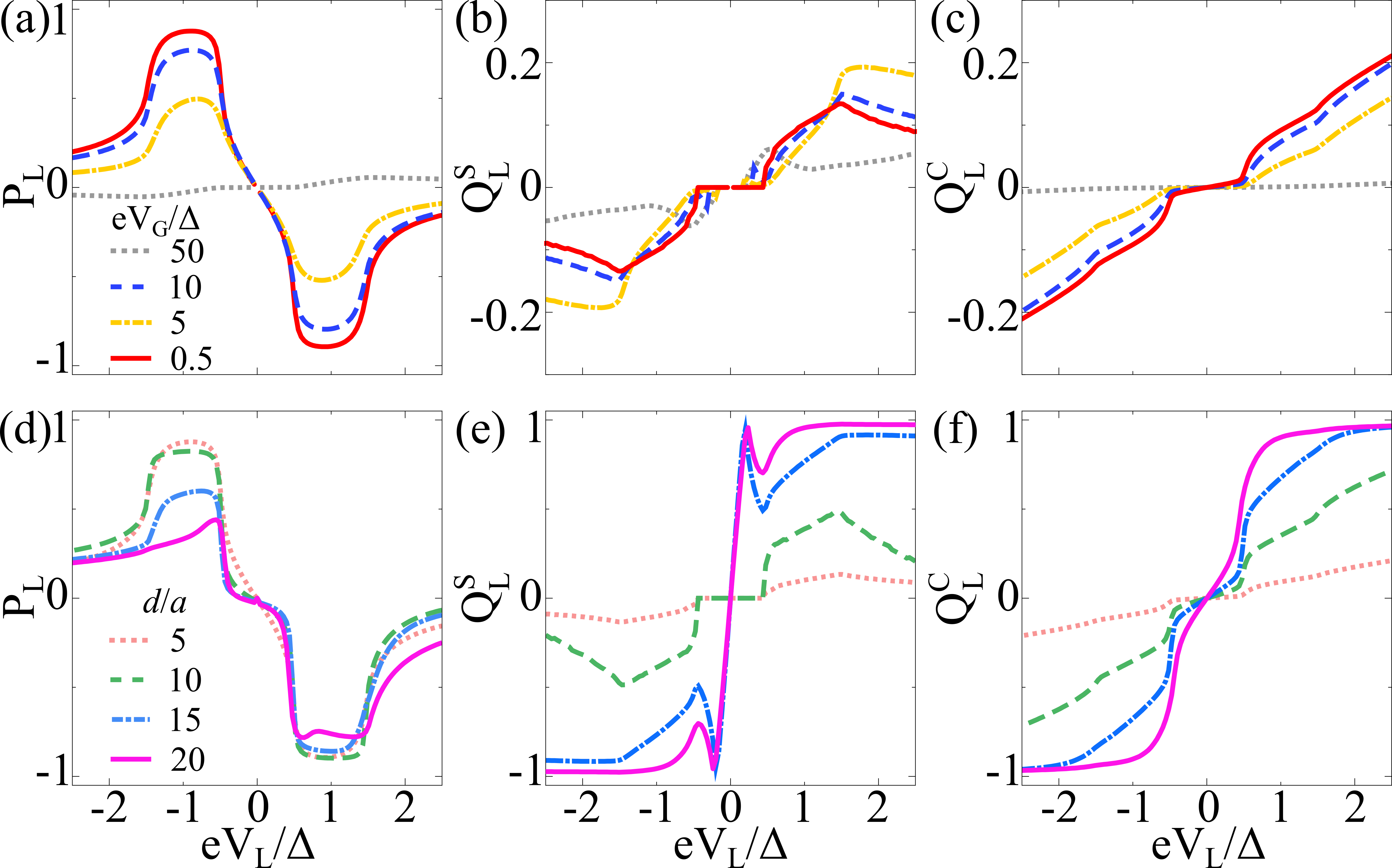}
\caption{
\textbf{Gate-controlled local spin and charge nonreciprocity.}
(a-c) Spin polarization $P_{\text{L}}$ and local spin (charge) quality factors $Q_{\rm L}^{\rm S(C)}$ as functions of $eV_{\rm L}$ for different gate voltages $eV_{\rm G}$ at fixed short length $d=5a$ in a junction having $d_{x^{2}-y^{2}}$-wave altermagnetism. (d-f) The same as in (a-c) but for different values of $d$ at fixed gate voltage  $eV_{\rm G}=0.5\Delta$.
Parameters: The same as in Fig.\,\ref{figure2}.  
}
\label{figure3}
\end{figure}

We start by inspecting in Fig.\,\ref{figure3} the spin polarization $P_{\text{L}}$ and the quality factors $Q_{\text{L}}^{\gamma}$ as functions of $eV_{\rm L}$ for distinct values of $eV_{\rm G}$ and length of the N$_{2}$ region $d$. 
By tuning the gate voltage $eV_{\rm G}$ from high to values of the order of $\Delta$ at fixed short $d$ as in Fig.\,\ref{figure2}(h), the spin polarization $P_{\text{L}}$ develops greatly enhanced values for $eV_{\rm L}$ in the partially gapped regime, see Fig.\,\ref{figure3}(a). 
This enhancement originates from the momentum filtering effect illustrated in Fig.\,\ref{figure1}(b-d), and discussed together with Figs.\,\ref{figure2}(d,h) in the previous section. 
Furthermore, the spin and charge nonreciprocal quality factos $Q_{\text{L}}^{\text{S}}$ and $Q_{\text{L}}^{\text{C}}$ exhibit nonzero values and depend on $eV_{\rm G}$, especially at large bias voltages, as shown in Figs.\,\ref{figure3}(b,c). 
This nonreciprocal behavior follows from the asymmetric conductances (currents) with respect to zero energy (zero bias) shown in Fig.\,\ref{figure2}(g,h), and can be understood from the schematics of Fig.\,\ref{figure1}(b-d).
In the N$_2$ region with $eV_{\rm G}\sim\Delta$, the equal-energy contour for $E>0$ is significantly larger than that for $E<0$, allowing more transverse modes to contribute under positive bias [Fig.\,\ref{figure1}(c,d)]. 
Consequently, together with the spin polarization caused by the mometum-filtering effect, the current for positive bias is dominated by spin-$\downarrow$ electrons, whereas under negative bias, fewer transverse modes contribute and the current is dominated by spin-$\uparrow$ electrons. Therefore, a larger spin-$\downarrow$ polarized current flows for positive bias, while a smaller spin-$\uparrow$ polarized current appears for negative bias. 
As a result, the nonzero difference between positive and negative charge currents leads to charge nonreciprocity, while the corresponding directional difference for the spin current produces spin nonreciprocity, as shown in Figs.\,\ref{figure3}(b,c). 
Note, however, that Figs.\,\ref{figure3}(b,c) show quality factors of the order of  $Q_{\text{L}}^{\text{C/S}}\sim0.2$, which occurs because the difference between the currents under opposite biases is moderate [Fig.\,\ref{figure2}(h)].

Extending the length of the gated region can further enhance the nonreciprocal effect, although at the cost of reducing the spin polarization, as shown in Figs.\,\ref{figure3}(d-f). 
In fact, when the length of the N$_2$ region increases, the spin polarization of the negative-bias current is strongly reduced, while the positive-bias current remains highly spin polarized with $P_{\text{L}}>0.5$, see Figs.\,\ref{figure3}(d). 
Importantly, both the spin and charge nonreciprocal efficiencies can approach $Q_{\text{L}}^{\text{C/S}}=\pm1$, indicating nearly perfect nonreciprocity. 
Unlike the high spin polarization that appears only in the partially gapped regime [Fig.\,\ref{figure3}(d)], the directional-selective effect occurs over a wider bias range when $eV_{\text{L}}>eV_{\rm G}$ and the N$_2$ region is sufficiently long [Fig.\,\ref{figure3}(e,f)]. 
The enhancement of the nonreciprocity can be understood from the schematic spectrum in Fig.\,\ref{figure1}(b). 
When the energy of the incident electron lies below the band bottom of the gated region, i.e., $E<-eV_{\rm G}<0$, quantum tunneling occurs through evanescent modes in the N$_2$ region. 
From the quantum scattering picture, the transmission probability scales as $e^{-\kappa d}$ with $\kappa=\sqrt{2m(eV_G-E)}/\hbar$, which decays exponentially with the length $d$ of the gated region. 
Therefore, the current is strongly suppressed when the bias enters this regime. 
In contrast, for positive bias, propagating states are always available in the N$_2$ region, so that a finite current persists. Hence, increasing the length of the gated region strongly suppresses the negative-bias spin-$\uparrow$ current, while the positive-bias current remains nearly unaffected.
Since the spin-$\downarrow$ current is already small, the two spin contributions become comparable under negative bias, leading to a reduction of the spin polarization [Fig.\,\ref{figure3}(d)]. 
At the same time, the strong suppression of the negative-bias current enhances both the spin and charge nonreciprocal quality factors [Fig.\,\ref{figure3}(e,f)].

\begin{figure}[t]
\centering \includegraphics[width=0.49\textwidth]{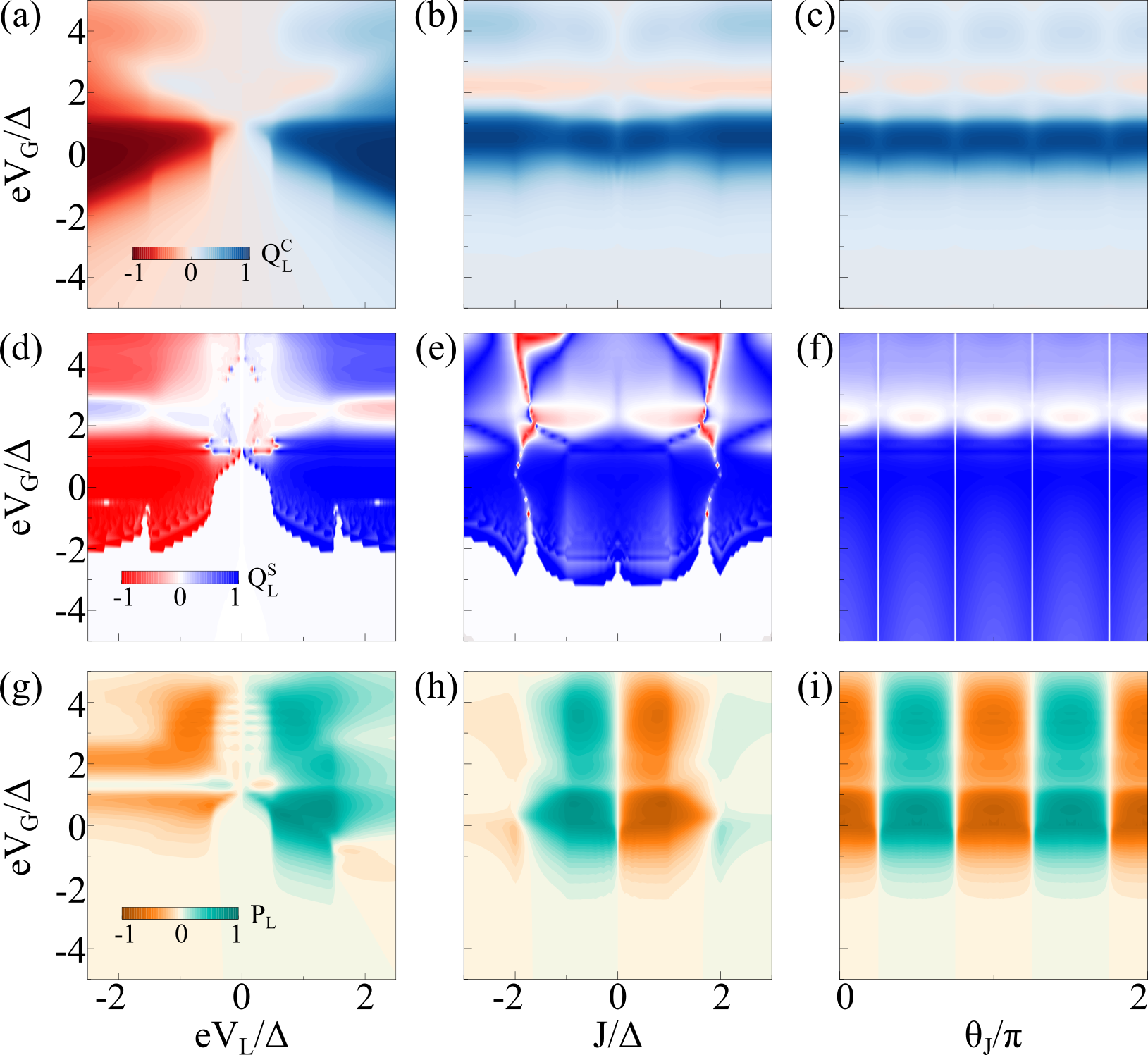}
\caption{
\textbf{Robustness of local spin and charge nonreciprocity.}
(a-c) Charge quality factor  as  a function of   $eV_{\rm G}$ and $eV_{\rm L}$ (a),  $J$ (b), and $\theta_{\rm J}$ (c). (d-f) and (g-i) are the same as (a-c) but for the spin quality factor   and spin polarization, respectively.  Parameters: same as in Fig.\,\ref{figure3} with $d=20a$; $eV_{\rm L}=\Delta$, $\theta_J=0$, and $J=0.5\Delta$ unless variable.
}
\label{figure4}
\end{figure}

To further inspect the robustness of the gate-controlled charge and spin nonreciprocity, in Fig.\,\ref{figure4} we present the quality factors $Q^{\text{C(S)}}_{\rm L}$ and spin polarization $P_{\rm L}$ as functions of the gate voltage $eV_{\rm G}$, bias $eV_{\rm L}$,   altermagnetic strength $J$, and altermagnetic orientation $\theta_J$, all for a $N_{2}$ region of length $d=20a$. The charge quality factor $Q^{\text{C}}_{\rm L}$ demonstrates that its polarity   depends on the bias $V_{\rm L}$ direction and reaches a nearly perfect value with $Q^{\text{C}}_{\rm L}\approx \pm1$ when $eV_{\rm G}$is comparable to the superconducting gap, see Fig.\,\ref{figure4}(a). 
Moreover, as long as the gate voltage is within this range, the high $Q^{\text{C}}_{\rm L}$ remains in a wide range of bias $eV_{\rm L}$ and is roughly insensitive to the altermagnetic strength $J$, and orientation $\theta_J$, provided the bias is fixed at $eV_{\rm L}/\Delta=1$, see Fig.\,\ref{figure4}(b,c). A similarly nearly perfect spin nonreciprocal behaviour with $Q^{\text{S}}_{\rm L}=\pm1$ appears when $eV_{\rm G}$ is comparable to the superconducting gap, which is also featured by a bias-direction-dependent polarity, see Fig.\,\ref{figure4}(d).
We note that, for low gate voltage $eV_{\rm G}/\Delta \ll 0$, however, $Q^{\text{S}}_{\rm L}$ becomes ill-defined because of the negligible current.
Notably, $Q^{\text{S}}_{\rm L}$ exhibits a strong dependence on the altermagnetic strength $J$ and and can even reverse its polarity [Fig.\,\ref{figure4}(e)].
Specifically, for a fixed $eV_{\rm G}$, enhancing the altermagnetic strength $J$ also reverses the spin nonreciprocal polarity at around $J/\Delta\sim\pm1$.
We attribute this polarity reversal to the switching of spin-polarized states in the superconducting AM. 
For a fixed $eV_{\rm G}$, we have confirmed that as $J$ increases, the energy of the spin-$\uparrow$ sector of the spectrum [blue curve in Fig.\,\ref{figure1}(b)] shifts upward, while the spin-$\downarrow$ sector shifts downward [red curve in Fig.\,\ref{figure1}(b)]. Consequently, the equal-energy contours shown in Fig.\,\ref{figure1}(c,d) interchange, reversing the spin polarization and thereby changing the polarity of $Q^{\text{S}}_{\rm L}$. 
This polarity reversal signals the exchange of the equal-energy contours in the altermagnet and is related to the emergence of the Bogoliubov-Fermi surface in superconducting AMs \cite{Lu2025Engineering,fukaya2026CFB}. 
Futhermore,, the polarity reversal also occurs by changing the gate voltage $eV_{\rm G}$ [Fig.\,\ref{figure4}(e)]; also, while $Q^{\text{S}}_{\rm L}$ seems almost insensitive to the altermagnetic orientation $\theta_{\rm J}$, it develops an oscillatory pattern with a polarity reversal at large $eV_{\rm G}$ [Fig.\,\ref{figure4}(f)].  
This gate-voltage-induced polarity reversal might be attributed to the gated-controlled resonant tunneling process that occurs in the confined N$_2$ region \cite{Fu2025All}.

A similar behavior is exhibited by the spin polarization $P_{\rm L}$ in Figs.\,\ref{figure4}(g-i), which demonstrate a large spin polarization as $eV_{\rm G}\sim\Delta$ and sign reversal as $J$ increases due to the switching of spin-polarized states in the superconducting AM,  see Fig.\,\ref{figure4}(g,h). An intriguing behavior of  $P_{\rm L}$ is that it develops an oscillatory behavior as a function of the altermagnetic orientation $\theta_{\rm J}$. 
Specifically, $P_{\rm L}=-1$ ($P_{\rm L}=+1$) occurs for $\theta_J=n\pi$ [$\theta_J=(2n+1)\pi/2$] when $eV_{\rm G}\sim\Delta$, corresponding to the directions of maximal spin splitting of the AM Fermi surface.
In contrast, $P_{\rm L}=0$ occurs at $\theta_J=(2n+1)\pi/4$, where transport occurs along the spin-degenerate nodes of the AM.
The $\theta_{\rm J}$ dependence of $P_{\rm L}$, therefore, provides strong signatures that allow to identify the type of altermagnetism.

Before ending, we emphasize that, despite the presence of a spin-split Fermi surface in superconducting AM hybrids, the absence of gating leads to vanishing spin-polarized transport due to the compensation between different spin channels.
Gating $N_{2}$ induces a momentum filtering effect, such that only modes with momenta matching the spin-polarized states in the superconducting AM contribute to transport, thereby enabling a finite spin-polarized current.
The gating further induces a nonreciprocity in the local current, which, combined with the momentum-selective spin filtering, gives rise to both spin and charge quality factors reaching nearly perfect values.
Beyond the $d$-wave altermagnetism considered here, the spin and charge nonreciprocities are expected to extend to other even-parity unconventional magnets, such as $g$-wave and $i$-wave systems. 
This follows from the presence of analogous spin-split spectra in these systems when proximitized by conventional $s$-wave superconductivity \cite{Yuri2025Superconducting,Fu2025Floquet,Fu2025Light}.
Consequently, the same gate-controlled momentum-filtering mechanism illustrated in Fig.\,\ref{figure1} can be applied to a broader class of even-parity unconventional magnets. 
In contrast, for odd-parity magnets such as $p$- and $f$-wave systems, although spin-split equal-energy contours remain, the spin-dependent energy shift in the spectrum [Fig.\,\ref{figure1}(b)] is absent. 
As a result, spin-polarized transport is suppressed, while we anticipate that nonreciprocal charge current can still arise by gating.

\subsection*{Nonlocal spin and charge nonreciprocity}
\begin{figure}[t]
\centering \includegraphics[width=0.49\textwidth]{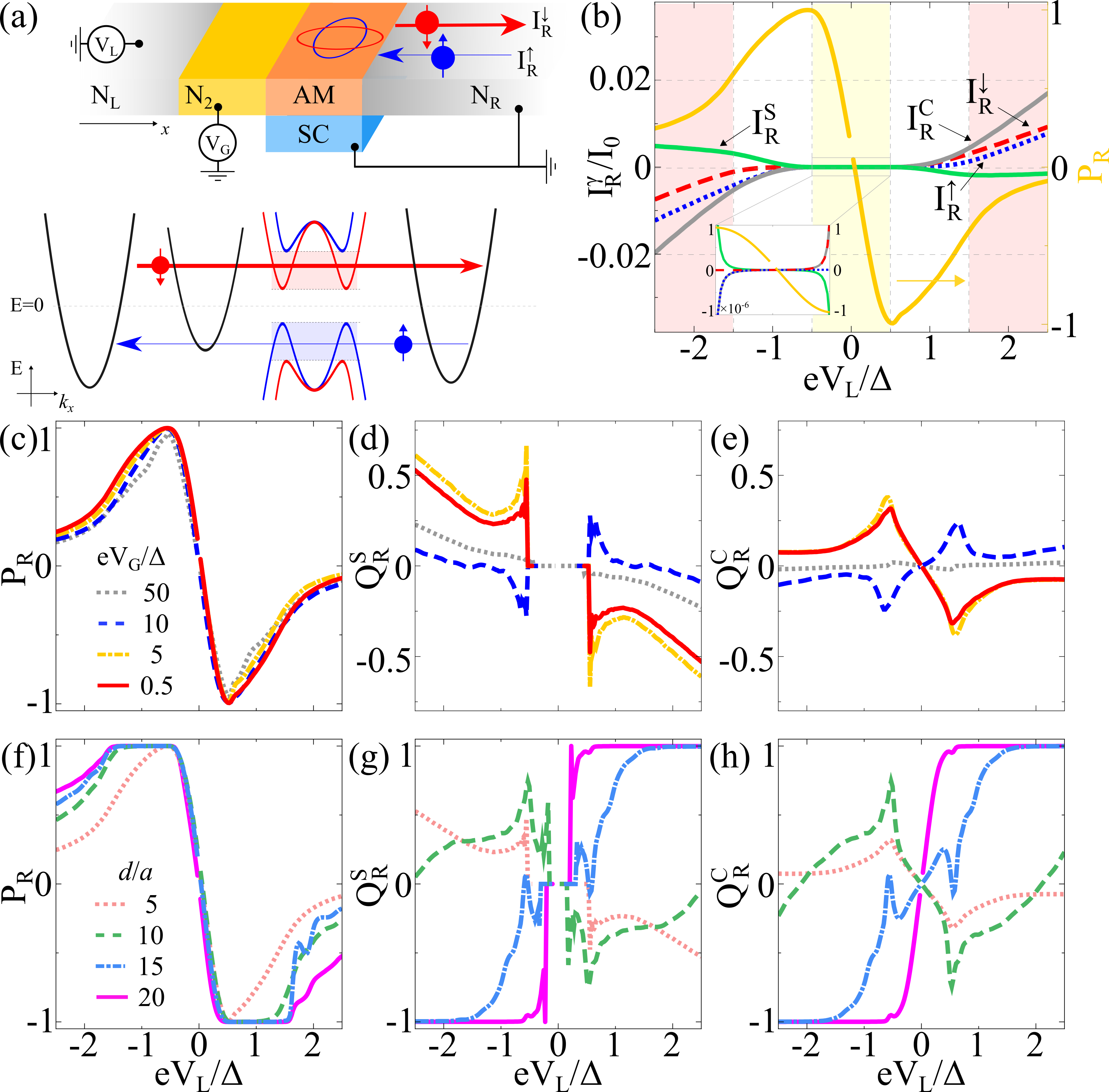}
\caption{
\textbf{Nonlocal spin and charge nonreciprocities.}
(a) Top: Schematic illustration of superconducting AM junction having $d_{x^{2}-y^{2}}$-wave altermagnetism and hosting nonlocal spin and charge currents, with the nonlocal up and down currents ($I_{\rm R}^{\sigma}$) indicated by horizontal red and blue arrows. 
The left lead is subjected to a bias voltage $V_{\rm L}$, while the finite region $N_{2}$ is gated by $V_{\rm G}$. 
The superconducting AM is of finite length $d_{\rm sc}$. 
Bottom: Energy dispersions in each region of the junction, with the shaded regions indicating the partially gapped energy windows in the superconducting AM.  
(b) Non-local current as a function of $eV_{\rm L}$, while the spin polarization $P_{\rm R}(eV_{\rm L})$ is shown in the right $y$ axis. 
The yellow and red shaded areas indicate gapped and fully gapless regions in the superconducting AM, respectively, while the white region between them corresponds to the partially gapped energy windows. 
(c-e) Spin polarization, spin, and charge quality factor as functions of $eV_{\rm L}$ for different gate voltages at fixed $d=5a$.
(f-h) The same as in (c-e) but for different $d$ at fixed $eV_{\rm G}=0.5\Delta$.
Parameters: $d_{\text{sc}}=500a$, while the rest is the same as in Fig.\,\ref{figure2}.
}
\label{figure5}
\end{figure}

The gate-induced momentum filtering mechanism discussed in the previous section also enables nonreciprocal nonlocal charge and spin currents. 
To study these nonlocal nonreciprocities, we consider the junction shown in Fig.\,\ref{figure5}(a), where we attach a normal metallic lead (N$_{\text{R}}$) on the right side of the superconducting AM of Fig.\,\ref{figure1}(a). 
Then, the resulting current in N$_{\text{R}}$ depends nonlocally on the bias $eV_{\text{L}}$ applied in N$_{\text{L}}$, and is obtained as \cite{Anantram1996Current,den1996Transport,Pierattelli2025Delta}
\begin{equation}
I_{\text{R}}^{\sigma}(eV_{\text{L}})
=\frac{1}{e}\int dE\, G_{\text{R}}^{\sigma}(E)
\left[f_{\text{L}}(E,eV_{\text{L}})-f_{\text{R}}(E)\right].
\label{eqIReV}
\end{equation}
Here, $f_{\text{L}}(E,eV_{\text{L}})$ and $f_{\text{R}}(E)=f_{\text{S}}(E)$ are the Fermi-Dirac distributions in N$_{\text{L}}$ and N$_{\text{R}}$, respectively, with bias $eV_{\text{L}}$ applied to N$_{\text{L}}$ and N$_{\text{R}}$ grounded.
As before, $G_{\text{R}}^{\sigma}(E)$ is the spectral nonlocal conductance given by
\begin{equation}
G_{\text{R}}^{\sigma}(E)
= G_{0}\sum_{k_{y}} g_{\text{R}}^{\sigma}(E,k_{y}),
\label{eqGGnl}
\end{equation}
with $G_{0}=e^{2}W/(2\pi\hbar)$ and
\begin{equation}
g_{\text{R}}^{\sigma}(E,k_{y})
= T_{e}^{\sigma}(E,k_{y}) - T_{h}^{\sigma}(E,k_{y}).
\label{eqgnl}
\end{equation}
where $T_{e}^{\sigma}$ ($T_{h}^{\sigma}$) represents the spin-resolved electron-electron cotunneling (crossed Andreev reflection) probability for an electron incident from N$_{\text{L}}$ with energy $E$ and transverse momentum $k_{y}$ transmitted as an electron (hole) into N$_{\text{R}}$. Thus, the nonlocal current in Eq.\,(\ref{eqIReV}) results from the competition between electron and hole transmission processes, and can   be either positive or negative depending on the dominant transport channel.
A positive (negative)   $I_{\text{R}}^{\sigma}(eV_{\text{L}})$ corresponds to current flowing from the superconducting AM region into N$_{\text{R}}$ (from N$_{\text{R}}$ into the superconducting AM region).  Thus, the current between N$_{\text{R}}$ and the superconducting AM region is controlled by the bias applied to the left lead N$_{\text{L}}$, hence unveiling its nonlocal character.  Following the quantities defined for the local currents  [see Eq.\,(\ref{eq_iv}) and Methods Section], Eq.\,(\ref{eqIReV}) allows us to define nonlocal transport observables, including the nonlocal charge (C) and spin (S) conductances: 
$G_{\text{R}}^{\text{C(S)}} = G_{\text{R}}^{\uparrow} \pm G_{\text{R}}^{\downarrow},$
and the corresponding nonlocal currents
$I_{\text{R}}^{\text{C(S)}} = I_{\text{R}}^{\uparrow} \pm I_{\text{R}}^{\downarrow}.$
Furthermore, the nonlocal transport is characterized by the spin polarization $P_{\text{R}}(eV_{\text{L}})$ and the charge (spin) nonreciprocal quality factor $Q_{\text{R}}^{\text{C(S)}}(eV_{\text{L}})$ in N$_{\text{R}}$, defined  as in Eqs.\,(\ref{eqP}) and (\ref{eqQ}) but with the currents given by Eq.\,(\ref{eqIReV}).

With the above considerations, we obtain the nonlocal spin and charge currents and show them in Fig.\,\ref{figure5}(b) as functions of $eV_{\rm L}$ at $eV_{\text{G}}=\Delta/2$; to complement, we also present the spin-resolved components $I_{\text{R}}^{\sigma}$ and the current spin polarization $P_{\text{R}}$. 
Although the nonlocal currents are weak, the spin-resolved nonlocal currents develop distinct values $I_{\text{R}}^{\uparrow}\neq I_{\text{R}}^{\downarrow}$. 
This gives rise to a finite spin current $I_{\text{R}}^{\text{S}}$ and thereby to a finite spin polarization $P_{\text{R}}$ within the gapped regime $|eV_{\text{L}}|<\Delta_{-}$ shown in the inset of Fig.\,\ref{figure5}(b).
Remarkably, at $|eV_{\text{L}}|=\Delta_{-}$, one spin component is almost negligible while the other remains measurable, resulting in perfect spin polarization $P_{\text{R}}(\pm\Delta_{-})=\mp 1$. 
This can be attributed to the intrinsic spin-split Fermi surface of the superconducting AM, instead of the gating effect.
More specifically, at $eV_{\text{L}}=\Delta_{-}$, only the corresponding transverse modes contribute to transport since $\Delta_{-}$ occurs at $\theta_{k}=0$ ($\theta_{k}=\pm\pi/2$) for the spin-$\downarrow$ (spin-$\uparrow$) subband [see Fig.\,\ref{figure1}(b-d)].
However, the spin-$\uparrow$ modes at $\theta_{k}=\pm\pi/2$ propagate parallel to the interface and therefore do not contribute to transmission, leaving only the spin-$\downarrow$ channel active and yielding $P_{\text{R}}(+\Delta_{-})=-1$. 
For $eV_{\text{L}}=-\Delta_{-}$, the situation is reversed, giving $P_{\text{R}}(-\Delta_{-})=+1$. 
Thus, this perfect nonlocal spin polarization is almost \textit{independent} of the gate-induced momentum filtering, as confirmed in Fig.\,\ref{figure5}(c). 
We have verified that this occurs even without the N$_2$ region, but then $P_{\text{R}}=\pm1$ is achieved at one point.
For $|eV_{\text{L}}|>\Delta_{-}$, the distinction between the spin-resolved nonlocal currents remains, but their difference reduces because both spin channels are engaged in the transport, resulting in decreasing spin polarization [Fig.\,\ref{figure5}(b)].
As the spin polarization is almost independent to the gate voltage, the presence of N$_2$ becomes more apparent when increasing its length in Fig.\,\ref{figure4}(f) at small $eV_{\text{G}}$. 
In fact, increasing the length of the N$_2$ region stabilizes the spin polarization, leading to $P_{\text{R}}(eV_{\text{L}})=\mp 1$ over a range within the partially gapped regime $\Delta_{-}<|eV_{\text{L}}|<\Delta_{+}$, as shown in Fig.\,\ref{figure4}(f). 
This robust and \textit{perfect} spin polarization of the nonlocal currents is absent in their local counterparts, see Fig.\,\ref{figure3}(d,h) for comparison.

Although gating the $N_{2}$ region weakly affects the spin polarization, it does play a crucial role in the nonlocal spin and charge nonreciprocities. 
We demonstrate this in Figs.\,\ref{figure4}(d,e) by presenting $Q_{\text{R}}^{\text{C(S)}}(eV_{\text{L}})$ for distinct values of $eV_{\rm G}$ at fixed short $d$, while in Figs.\,\ref{figure4}(g,h) we instead vary $d$ at fixed $eV_{\rm G}$ comparable to $\Delta$. The first observation is that overall $Q_{\text{R}}^{\text{C(S)}}(eV_{\text{L}})\neq0$, implying that both nonlocal spin and charge currents exhibit a clear nonreciprocity. By tuning the gate voltage $V_{\rm G}$ and the length $d$ of the N$_2$ region, the spin quality factor is significantly enhanced once the bias exceeds the lower gap edge $|eV_{\text{L}}|>\Delta_{-}$, as seen in Figs.\,\ref{figure4}(d,g). In this regard, while lowering $eV_{\rm G}$ enhances $Q_{\text{R}}^{\text{S}}(eV_{\text{L}})$ [Fig.\,\ref{figure4}(d)], increasing $d$ notably leads to perfect spin quality factor $Q_{\text{R}}^{\text{S}}(eV_{\text{L}})=\pm1$ [Fig.\,\ref{figure4}(g)]. The increase of $d$ also induces an oscillating structure before reaching constant perfect values; this structure comes from confinement effects \cite{PhysRevB.91.024514,PhysRevB.104.L020501,PhysRevB.104.134507,Fu2025All} as a result of the finite length and lead to an oscillating conductance pattern. Furthermore, $eV_{\rm G}$ and $d$ also permit control of the polarity of the spin nonreciprocity; this originates from the competition between electron cotunneling and crossed Andreev reflection processes, allowing the net current to change sign depending on the dominant process. 
Similar perfect nonreciprocal behavior and controllable polarity are observed for the  charge current in Figs.\,\ref{figure4}(e,g). 
Compared to the local current shown in Figs.\,\ref{figure2} and \ref{figure3}, the nonlocal transport directly reflects the intrinsic spin-split Fermi surface of the  AM, resulting in robust and perfect spin polarization. At the same time, both charge and spin nonreciprocal effects remain strong and can reach the ideal limit under suitable conditions for the gated region.


\section*{Conclusions}
In conclusion, we have demonstrated that perfect spin and charge nonreciprocity is achieved in superconducting altermagnet heterostructures due to the interplay between altermagnetism and gate-controlled momentum filtering.
Focusing on $d$-wave altermagnets proximitized by conventional $s$-wave superconductivity, we showed that a finite gated normal region acts as a momentum filter that directionally selects transverse modes matching the momentum-dependent spin-split states in the superconducting altermagnet.
As a result, local and nonlocal spin currents exhibit nearly perfect and perfect, respectively, with highly tunable polarity, by means of tuning gate voltages and varying the length of the finite region. 
Moreover, we find that local and nonlocal charge currents also develop a nonreciprocal behavior, with quality factors approaching perfect values. 
Beyond nonreciprocity, the spin and charge currents are sensitive to variations of the altermagnetic field, a functional dependence that helps identify the strength and type of the altermagnetism.
Given that similar device architectures have been realized in ferromagnet-superconductor hybrids \cite{Strambini2022Superconducting}, our proposal is experimentally accessible.
Candidate platforms include $d$-wave altermagnets such as RuO$_2$ films \cite{Howzen2026Transport}, as well as systems hosting even-parity unconventional magnetism proximitized by $s$-wave superconductors \cite{Yuri2025Superconducting,Fu2025Floquet,Fu2025Light}, including In-MnTe \cite{Kazmin2025Andreev}, Co$_{1/4}$NbSe$_2$ \cite{DeVita2025Optical}, In-CrSb \cite{Esin2026Josephson}, and noncollinear altermagnets such as Mn$_3$Pt/Nb heterostructures \cite{Schrade2026AltermagneticTheory,Sachin2026AltermagneticExperiments}.
Our results establish a mechanism for generating nonreciprocal superconducting spin transport without net magnetization or external magnetic fields, providing a practical route toward electrically controlled superconducting spintronic devices based on altermagnets.


\section*{Methods}
In this section, we present the calculation of the spin-resolved local and nonlocal currents defined in Eqs.\,(\ref{eq_iv}) and (\ref{eqIReV}), respectively.
These currents are obtained from the corresponding spectral conductances, Eqs.\,(\ref{eq_Gx}) and (\ref{eqGGnl}), which are expressed in terms of scattering coefficients.
For the local current, these include the spin-resolved normal and Andreev reflection probabilities $R_{\text{N}}^{\sigma}$ and $R_{\text{A}}^{\sigma}$, while for the nonlocal current they include electron--electron cotunneling and crossed Andreev reflection processes, characterized by $T_{e}^{\sigma}$ and $T_{h}^{\sigma}$.
The scattering coefficients are computed numerically using the lattice Green's-function method \cite{Anantram2008Modeling}.
In the following, we detail the implementation of this approach for the N$_{\text{L}}$-N$_2$-S junction shown in Fig.\,\ref{figure1}(a), starting from the construction of the tight-binding Hamiltonian.

\subsection*{Tight-binding Hamiltonian of the junction in Fig.\,\ref{figure1}(a)}
The N$_{\rm L}$-N$_2$-S junction is modulated by the Hamiltonian 
\begin{equation}
H_{\text{J}}=
\begin{pmatrix}
H_{\text{L}} & T_{\text{LN}} & 0 \\
T_{\text{LN}}^{\dagger} & H_{\text{N}} & T_{\text{NS}} \\
0 & T_{\text{NS}}^{\dagger} & H_{\text{S}}
\end{pmatrix},
\label{eq_hj}
\end{equation}
where $H_{\text{L}}$, $H_{\text{N}}$, and $H_{\text{S}}$ describe the left normal-metal lead (N$_\text{L}$), the central gated region (N$_2$), and the right lead of $d$-wave altermagnets with $s$-wave superconductivity, respectively. 
The off-diagonal blocks $T_{\text{LN}}$ and $T_{\text{NS}}$ denote the N$_\text{L}$-N$_2$ and N$_2$-S couplings, respectively.
We first construct the superconducting block $H_{\text{S}}$. 
Since the junction is oriented along the $x$ direction, translational symmetry along $y$ is preserved, and $k_y$ remains a good quantum number. 
Discretizing Eq.\,(\ref{eq_hqbdg}) along $x$ direction, we obtain
\begin{equation}
H_{\text{S}}=
\begin{pmatrix}
h_{0}(k_y) & h_{x}(k_y) & 0 & \cdots \\
h_{x}^{\dagger}(k_y) & h_{0}(k_y) & h_{x}(k_y) & \cdots \\
0 & h_{x}^{\dagger}(k_y) & \ddots & \ddots \\
\vdots & \vdots & \ddots & \ddots
\end{pmatrix},
\label{eq_hstb}
\end{equation}
which is an infinite-dimensional matrix 
with elements
\begin{eqnarray}
h_{0}(k_y) &=& (4t-\mu)\tau_z\sigma_0-\Delta\tau_y\sigma_y \nonumber \\
& &-2\!\left[t\tau_z\sigma_0-J\cos(2\theta_J)\tau_z\sigma_z\right]\!\cos(k_y),
\label{eq_h0} \\
h_{x}(k_y) &=& -t\tau_z\sigma_0-J\cos(2\theta_J)\tau_z\sigma_z \nonumber \\
& &+iJ\sin(2\theta_J)\sin(k_y).
\label{eq_hx}
\end{eqnarray}
Here $\sigma_j$ ($\tau_j$) is the $j$-th Pauli matrix in spin (Nambu) space, $t=\hbar^2/(2ma^2)$ is the nearest-neighbor hopping amplitude, and $a$ is the lattice constant. 
Throughout this work, we set $a=1$ and $t=1$ as the units of length and energy.
Eqs.\,(\ref{eq_h0}) and (\ref{eq_hx}) are obtained by Fourier transforming Eq.\,(\ref{eq_hqbdg}) after substituting
$ak_i\!\rightarrow\!\sin ak_i$ and $(ak_i)^2\!\rightarrow\!2(1-\cos ak_i)$ ($i=x,y$).
The normal-metal Hamiltonian $H_{\text{L}}$ follows from Eq.\,(\ref{eq_hstb}) by setting $J=\Delta=0$, which gives
\begin{equation}
H_{\text{L}}=
\begin{pmatrix}
\ddots & \ddots & \vdots & \vdots \\
\ddots & \ddots & h_{\text{L},x}(k_y) & 0 \\
\cdots & h_{\text{L},x}^{\dagger}(k_y) & h_{\text{L},0}(k_y) & h_{\text{L},x}(k_y) \\
\cdots & 0 & h_{\text{L},x}^{\dagger}(k_y) & h_{\text{L},0}(k_y)
\end{pmatrix},
\label{eq_hntb}
\end{equation}
with
\begin{eqnarray}
h_{\text{L},0}(k_y) &=& (4t-\mu)\tau_z\sigma_0-2t\cos(k_y)\tau_z\sigma_0, \\
h_{\text{L},x}(k_y) &=& -t\tau_z\sigma_0.
\end{eqnarray}
We assume an identical chemical potential $\mu$ and hopping strength in the normal and superconducting leads. 
Then, the central scattering region of length $d$ is described by
\begin{equation}
H_{\text{N}}=
\begin{pmatrix}
h_{\text{N},0}(k_y) & h_{\text{N},x}(k_y) & 0 & \cdots \\
h_{\text{N},x}^{\dagger}(k_y) & \ddots & \ddots & 0 \\
0 & \ddots & h_{\text{N},0}(k_y) & h_{\text{N},x}(k_y) \\
\cdots & 0 & h_{\text{N},x}^{\dagger}(k_y) & h_{\text{N},0}(k_y)
\end{pmatrix},
\label{eq_Hbtb}
\end{equation}
with 
\begin{eqnarray}
h_{\text{N},0}(k_y) &=& (4t-eV_{\text{G}})\tau_z\sigma_0-2t\cos(k_y)\tau_z\sigma_0, \\
h_{\text{N},x}(k_y) &=& h_{\text{L},x}(k_y).
\end{eqnarray}
where the gate voltage $eV_{\text{G}}$ plays the same role as the chemical potential, shifting the bands.
The N$_\text{L}$-N$_2$ coupling in modulated by
\begin{equation}
T_{\text{LN}}=(t_c/t)\,
\begin{pmatrix}
\vdots & \vdots & \cdots \\
0 & 0 & \cdots \\
h_{\text{L},x}(k_y) & 0 & \cdots
\end{pmatrix},
\end{equation}
and the N$_2$-S coupling is
\begin{equation}
T_{\text{NS}}=(t_c/t)\,
\begin{pmatrix}
\vdots & \vdots & \cdots \\
0 & 0 & \cdots \\
h_{\text{N},x}(k_y) & 0 & \cdots
\end{pmatrix},
\label{eq_tns}
\end{equation}
where $t_c$ is the interface coupling strength. 
The limits $t_c/t=1$ and $t_c/t\ll1$ correspond to ballistic and tunneling transport regimes, respectively.
In Eq.\,\ref{eq_hj}, $H_{\text{L}}$ and $H_{\text{S}}$ describe semi-infinite leads, whereas $H_{\text{N}}$ has a finite dimension $4d\times4d$, where the factor $4$ accounts for the spin and Nambu degrees of freedom.
Moreover, we assume that the hopping amplitude and chemical potential are identical in N$_\text{L}$ and S regions. Consequently, scattering arises solely from 
(i) Fermi-surface mismatch in N$_2$ induced by gating $eV_{\text{G}}$, and 
(ii) the interface transparency controlled by the tunneling parameter $t_c/t$, where $t_c/t=1$ ($t_c/t\ll1$) corresponds to the ballistic (tunneling) transport regime.

\subsection*{Scattering coefficients}
After constructing the junction Hamiltonian [Eq.\,(\ref{eq_hj})], the scattering matrix can be is obtained using the lattice Green's function technique \cite{Datta1997Electronic}.
Employing the Fisher-Lee relation \cite{Fisher1981Relation}, the scattering matrix of the junction reads \cite{Xu2022Electrically, Zhang2017Quantum, Fu2020Transport,Fu2022, Fu2022b}
\begin{widetext}
\begin{equation}
S_{q,p}^{(\sigma ,\alpha ),(\sigma ^{\prime },\beta )}(E,k_{y})
=-\delta_{p,q}\delta_{\alpha ,\beta }\delta_{\sigma ,\sigma ^{\prime }}
+i[\Gamma_{q}^{\sigma ,\alpha }(k_{y})]^{1/2}
G_{q,p}^{(\sigma ,\alpha ),(\sigma ^{\prime },\beta )}(k_{y})
[\Gamma_{p}^{\sigma ^{\prime },\beta }(k_{y})]^{1/2}.
\label{FisherLee}
\end{equation}
\end{widetext}
Here $\alpha,\beta\in\{e,h\}$ label electron and hole sectors with spin
$\sigma,\sigma'$, and $p,q\in\{\text{L},\text{S}\}$ denote the N$_\text{L}$ and
superconducting leads, respectively.
The matrix element
$S_{q,p}^{(\sigma ,\alpha ),(\sigma ^{\prime },\beta )}(E,k_{y})$
represents the scattering amplitude for an incident quasiparticle of type
$(\sigma',\beta)$ from lead $p$ being scattered into $(\sigma,\alpha)$ in
lead $q$.
The linewidth function is defined as
$\Gamma_{p}(k_{y})=i[\Sigma_{p}^{r}(k_{y})-\Sigma_{p}^{a}(k_{y})]$,
where $\Sigma_{p}^{r/a}(k_{y})$ is the retarded/advanced self-energy due to
coupling to lead $p$, calculated numerically via the recursive method
\cite{LopezSancho1985Highly, LopezSancho1985Nonorthogonal, LopezSancho1984Quick}.
The retarded Green's function of the scattering region is
\begin{equation}
G^{r}(E,k_{y})
=[E-H_{\text{N}}-\Sigma_{\text{L}}^{r}(k_{y})-\Sigma_{\text{S}}^{r}(k_{y})]^{-1}.
\end{equation}
The normal- and Andreev-reflection probabilities entering the conductance
formulas in the main text are then obtained as
\begin{eqnarray}
R_{\text{N}}^{\sigma }(E,k_{y})
&=&\mathrm{Tr}\!\left[
S_{\text{L},\text{L}}^{(\sigma ,e),(\sigma ,e)\dagger }
S_{\text{L},\text{L}}^{(\sigma ,e),(\sigma ,e)}
\right], \label{eq_RN}
\\
R_{\text{A}}^{\sigma }(E,k_{y})
&=&\mathrm{Tr}\!\left[
S_{\text{L},\text{L}}^{(\bar{\sigma},h),(\sigma ,e)\dagger }
S_{\text{L},\text{L}}^{(\bar{\sigma},h),(\sigma ,e)}
\right], \label{eq_RA}
\end{eqnarray}
where $\bar{\sigma}=-\sigma$.
Eqs.\,(\ref{eq_RN}) and (\ref{eq_RA}) describe the probabilities of an incident electron reflected as a backscattering electron with equal spin and Andreev reflected into an opposite-spin hole, injecting Cooper pairs into the superconductors, respectively.
Then, by substituting Eqs.\,(\ref{eq_RN}) and (\ref{eq_RA}) into Eq.\,(\ref{eq_Ts}), the spin-resolved spectral conductance [Eq.\,(\ref{eq_Gx})], and subsequently the current [Eq.\,(\ref{eq_iv})] are obtained.

\subsection*{Calculation of the nonlocal current in Fig.\,\ref{figure4}(a)}

Next, we present the details of the nonlocal current flowing in the setup given in Fig.\,\ref{figure4}(a).
The nonlocal current defined in Eq.\,(\ref{eqIReV}) can be obtained following a procedure analogous to the above-mentionsed local case. 
The Hamiltonian describing the junction in Fig.\,\ref{figure4}(a) retains the block-diagonal structure of Eq.\,(\ref{eq_hj}), with two key modifications:
First, the superconducting region is described by a $4d_{\text{S}}\times 4d_{\text{S}}$ matrix, where $d_{\text{S}}$ denotes the length of the superconducting altermagnets. 
Second, an additional normal lead N$_{\text{R}}$ is attached to the right of the superconducting region, described by a semi-infinite tight-binding Hamiltonian identical to that of N$_{\text{L}}$ [Eq.\,(\ref{eq_hntb})].
Within the Fisher-Lee formalism [Eq.\,(\ref{FisherLee})], the lead indices are now extended to $p,q \in \{\text{L}, \text{R}\}$ to account for transport between the two normal leads. 
The spin-resolved electron-electron cotunneling and crossed Andreev reflection probabilities are then given by
\begin{eqnarray}
T_{e}^{\sigma}(E,k_{y})
&=&
\mathrm{Tr}\!\left[
S_{\text{R},\text{L}}^{(\sigma,e),(\sigma,e)\dagger}
S_{\text{R},\text{L}}^{(\sigma,e),(\sigma,e)}
\right], 
\label{eq_Te}
\\
T_{h}^{\sigma}(E,k_{y})
&=&
\mathrm{Tr}\!\left[
S_{\text{R},\text{L}}^{(\bar{\sigma},h),(\sigma,e)\dagger}
S_{\text{R},\text{L}}^{(\bar{\sigma},h),(\sigma,e)}
\right],
\label{eq_Th}
\end{eqnarray}
respectively.
Substituting Eqs.\,(\ref{eq_Te}) and (\ref{eq_Th}) into Eq.\,(\ref{eqgnl}), the nonlocal conductance is obtained, and the corresponding nonlocal current follows from Eq.\,(\ref{eqIReV}).

\section*{Data Availability}
The data supporting the findings of this study are available from the corresponding authors upon reasonable request.
   
\section*{Acknowledgements}
P. H. F. thanks the support from W. X..
P. H. F. and L. C. acknowledge financial support from the Fondazione Cariplo under the grant 2023-2594.
J.-F. L. acknowledges financial support from the National Natural Science Foundation of China (Grant No. 12174077). 
J. C. acknowledges financial support from the Swedish Research Council (Vetenskapsr{\aa}det Grant No. 2021-04121) and from the Olle Engkvist Foundation (Grant No. 243-1026).  

\section*{Author Contributions}
P.-H. F. and J. C. conceived the idea. 
P.-H. F. carried out the analytical and numerical calculations, made the figures, and prepared the manuscript.  
J.-F. L. provided calculation resources, valuable insights, and contributed to the analysis and interpretation of the results.  L. C. and J. C.  supervised the project, contributed to the interpretation of results, and helped write the final version of the manuscript.  All authors discussed the results and contributed to the final version of the manuscript.

\section*{Competing Interests}
The authors declare no competing interests.

\bibliography{biblio}

\end{document}